\documentclass[aps,prl,reprint,superscriptaddress]{revtex4-1}
\usepackage[utf8]{inputenc}

\usepackage{graphicx}  
\usepackage{xcolor}
\usepackage{bm}        

\usepackage{amssymb} 
\usepackage{amsmath}
\usepackage{amsthm}
\usepackage{amsfonts}
\usepackage{mathtools}
\usepackage{ragged2e}
\usepackage{braket}
\usepackage{comment}
\usepackage{tikz}
\usepackage{hyperref}

\begin{document}

\title{Better product formulas for quantum phase estimation}

\author{Kasra Hejazi}
\affiliation{Xanadu, Toronto, ON, M5G2C8, Canada}

\author{Jay Soni}
\affiliation{Xanadu, Toronto, ON, M5G2C8, Canada}

\author{Modjtaba Shokrian Zini}
\affiliation{Xanadu, Toronto, ON, M5G2C8, Canada}

\author{Juan Miguel Arrazola}
\affiliation{Xanadu, Toronto, ON, M5G2C8, Canada}

\begin{abstract}
Quantum phase estimation requires simulating the evolution of the Hamiltonian, for which product formulas are attractive due to their smaller qubit cost and ease of implementation. However, the estimation of the error incurred by product formulas is usually pessimistic and task-agnostic, which poses problems for assessing their performance in practice for problems of interest. In this work, we study the error of product formulas for the specific task of quantum energy estimation.
To this end, we employ the theory of Trotter error with a Magnus-based expansion of the effectively simulated Hamiltonian. The result is a generalization of previous energy estimation error analysis of gapped eigenstates to arbitrary order product formulas. As an application, we discover a 9-term second-order product formula with an energy estimation error that is quadratically better than Trotter-Suzuki.
Furthermore, by leveraging recent work on low-energy dynamics of product formulas, we provide tighter bounds for energy estimation error in the low-energy subspace. We show that for Hamiltonians with some locality and positivity properties, the cost can achieve up to a quadratic asymptotic speedup in terms of the target error. 
\end{abstract}

\maketitle

Time evolving a quantum system under a Hamiltonian, usually referred to as Hamiltonian simulation, is a ubiquitous subroutine in many quantum algorithms. A variety of methods have been proposed for performing Hamiltonian simulation \cite{lloyd1996universal,suzuki1991general,campbell2019random,berry2015simulating,low2017optimal,low2019hamiltonian,low2019well}, but product formulas remain one of the main contenders for implementation, especially on noisy or early fault-tolerant quantum computers. Determining the cost of implementing product formulas is usually based on an accurate analysis of the error induced in time evolution \cite{childs2021theory,childs2018toward}. As seen by the relatively few works in the literature that take specifics of the problems into account \cite{yi2022spectral,csahinouglu2021hamiltonian,hejazi2024better}, in most cases the error of the whole time evolution operator is assumed to represent the error for performing specific tasks of interest. This assumption generally results in a pessimistic overestimation of the cost of implementation~\cite{childs2021theory}. 

One prominent such task is energy estimation, carried out by techniques such as quantum phase estimation (QPE), wherein quantum Hamiltonian simulation is a core subroutine. Our focus in this work is on obtaining the error in the outcomes of QPE when using product formulas. 
We should note that other than the input/output structure, we will not be using anything related to the inner workings of QPE in our proofs, and thus our results hold for any energy estimation routine outputting an (approximate) energy of the input Hamiltonian.
We note prior work of Ref.~\cite{yi2022spectral}, arguing that whenever the standard first-order Trotter formula is used for QPE, the induced error in the outcomes is asymptotically smaller than what is expected from the general bounds for product formulas: the scaling of the step size with the QPE target error $\epsilon$ follows an $\mathcal{O}(\epsilon^{1/2})$ behavior rather than the expected $\mathcal{O}(\epsilon)$ scaling from the general bounds. 

We now discuss our contributions. First, we generalize previous results on QPE error \cite{yi2022spectral} to product formulas of arbitrary order; more precisely, for the error in energy estimation of eigenstates with a spectral gap, we obtain closed-form formulas as perturbative expansions in powers of the Trotter step-size; notably, the coefficients are expressed in terms of nested commutators of the terms of the Hamiltonian~\cite{childs2021theory}. 
Using this perturbative analysis, we introduce a general approach for obtaining custom-designed product formulas that are tailored to obtain asymptotically lower error in energy estimation. 
Based on that, for Hamiltonians of the form $H=A+B$, we discover new alternating product formulas of the form $ABABA$ outperforming the comparable second-order Trotter-Suzuki of the form $ABA$. Examples of such $H$ include electronic structure Hamiltonians and different types of spin chain models. We numerically verify our findings for a two-dimensional $3\times 4$ XY lattice Hamiltonian.

We also obtain an improved QPE error bound for energy estimation for low-lying states. We connect our analysis to that of \cite{hejazi2024better} on low-energy bounds for product formulas for a class of $k$-local Hamiltonians. When the error is written as a Taylor series in terms of the Trotter step-size, we show a tighter bound for the smaller order contributions: The coefficients grow only polylogarithmically in system size instead of polynomially, as predicted by standard bounds based on nested commutators. This can result in up to quadratically improving the asymptotic cost scaling  in terms of the target error.

\textit{Notation and terminology---}For a product formula implementation, an evolution $e^{-iHT}$ for a total time $T$ is generally broken into a number of Trotter steps $r$, each with length $s=T/r$. Product formulas approximating $e^{-iHs}$ require a Hamiltonian decomposition $H = \sum_{j=1}^M H_j$ to terms $H_j$ for which the evolution can be implemented in sublinear time (i.e.~fast-forwardable terms):
\begin{equation}\label{eq:general_product_formula}
    \mathcal{W}(s) = e^{-i a_qs H_{m_q}} \ldots e^{-i a_2s H_{m_2}} \, e^{-i a_1s H_{m_1}}.
\end{equation}
Here, $q$ is the number of such exponentials in the product formula $\mathcal{W}(s)$, $a_j$'s are the coefficients of the product formula, and $m_1,\ldots,m_q$ is a sequence of a (repeated) permutation of $1,\ldots,M$. The time evolution error for a single step $\varepsilon(s)$ is calculated and the total error, which is additive, follows as $r \varepsilon(s)$. 
The `order' of a product formula is defined based on the accumulated error in a single Trotter step: an order $p$ product formula has error $\varepsilon(s) = \mathcal{O}(s^{p+1})$. More precisely \cite{childs2021theory}:
\begin{equation}\label{eq:error_operator}
\begin{aligned}
    \left\| e^{-iHs}- \mathcal{W}(s) \right\| = \mathcal{O} \bigg( s^{p+1} \, \sum_{m=1}^M \sum_{\{m_i\} } f_{\{m_i\},m}  \\
    \times   \left\|  [H_{m_p}, \ldots ,[H_{m_1},H_m]\ldots]  \right\| \bigg),
\end{aligned}    
\end{equation}
where $m_i$'s run over all terms of the Hamiltonian, and $f_{\{m_i\},m}$ are $\mathcal{O}(1)$ positive numbers determined by the $a_j$'s. 

\section{Improved Magnus-based QPE Error Bound}
We follow the approach in \cite{childs2021theory} for obtaining the error for product formulas, which boils down  to expressing the error in terms of nested commutators of the Hamiltonian terms. 
Evolution for the total time requires $T/s$ steps, and the errors will accumulate resulting in an $\mathcal{O}(s^p)$ scaling. 
In general, the error in outcomes of a QPE that utilizes an order $p$ product formula has the same $\mathcal{O}(s^p)$ scaling.

We develop a more refined error analysis for energy estimation; based on the analysis done in \cite{yi2022spectral}, an effective Hamiltonian $\tilde{H}(s)$ can be defined that satisfies:
\begin{equation}\label{eq:def_Htilde}
    e^{-i\tilde{H}(s) s} = \mathcal{W}(s).
\end{equation}
The QPE algorithm starts with an initial state $\ket{\psi_{\text{in}}}$ and upon measurement yields $\ket{\tilde{\psi}(s)}$ and $\tilde{E}(s)$, an eigenstate and eigenvalue of $\tilde{H}(s)$, which can be viewed as approximations to an eigenstate $\ket{\psi}$ and eigenvalue $E$ of $H$, respectively. The error $\tilde{E}(s)-E$ can be interpreted as the deviation of the eigenvalues of $\tilde{H}(s)$ resulting from a perturbation on $H$. Hence, by perturbation theory in powers of $\tilde{H}-H$ to lowest order, the change in the effective energy reads:
\begin{equation}\label{eq:energy_shift_pert}
\begin{aligned}
    \delta E(s) = \tilde{E}(s)-E &= \langle \psi | (\tilde{H}(s) - H) |\psi\rangle \\
    &+ \text{[higher order terms]}.
\end{aligned}
\end{equation}
Building on \cite{yi2022spectral}, we present a more detailed perturbation theory analysis in the following. We note in passing that a nonvanishing gap between the eigenvalue of interest and other eigenvalues is assumed for all time step values between 0 and $s$. We present more details on this gap in the supplementary material, wherein we also treat the higher-order effects in the perturbation theory. Going forward, we drop $s$ from $\tilde{H}(s)$ and $\delta E(s)$ for notational simplicity.

We first need to derive the effective Hamiltonian. We use the multiplicative error representation in \cite{childs2021theory},  and rewrite the product formula as a time-ordered exponential:
\begin{equation}\label{eq:time_ordered_w}
    \mathcal{W}(s) = \mathcal{T} e^{-i \int_0^s  d\sigma \; \mathcal{F}(\sigma) },
\end{equation}
with $\mathcal{F}(\sigma) = H + \mathcal{E}(\sigma)$, where $\mathcal{E}(\sigma)$, called the exponent operator, consists of a sum of nested commutators of the terms of the Hamiltonian \cite{childs2021theory}. For a product formula of order $p$, this operator is expressed as a sum of nested commutators of length $p$ and above, where each nested commutator $[H_{m_l}, \ldots ,[H_{m_1},H_m]\ldots]$ of length $l$ is scaled by coefficients of order $s^l$ (see the supplementary material for the exact expression). 

In order to transform the time-ordered exponential representation in Eq.~\eqref{eq:time_ordered_w} to a single exponential containing $\tilde{H}$ as in Eq.~\eqref{eq:def_Htilde}, we use a Magnus expansion \cite{blanes2009magnus}. This is similar to the approach used in Ref.~\cite{tran2021faster} for a low-order calculation of an effective Hamiltonian. This allows us to calculate the effective Hamiltonian order by order as follows:
\begin{equation}
    \tilde{H} = \tilde{H}_1 + \tilde{H}_2 + \tilde{H}_3 + \ldots,
\end{equation}
and each of the terms defined as \cite{arnal2018general}:
\begin{equation}\label{eq:magnus_for_Htilde}
\begin{aligned}
    \tilde{H}_1 &= \frac{1}{s} \int_0^s d\sigma \; \mathcal{F}(\sigma), \\
    \tilde{H}_2 &= \frac{1}{2s} \int_0^s d\sigma_1 \int_0^{\sigma_1} d\sigma_2 \; [\mathcal{F}(\sigma_1) , \mathcal{F}(\sigma_2) ] \\
    \vdots \\
    \tilde{H}_n & = 
    \frac{1}{ns} \sum_\rho (-1)^{d_\rho} \frac{1}{{n-1 \choose d_\rho}} \int_0^s d\sigma_1 \ldots \int_0^{\sigma_{n-1}} d\sigma_n \; \\
    & \qquad [ \mathcal{F}(\sigma_{\rho(1)}) , \ldots [\mathcal{F}(\sigma_{\rho(n-1)}) , \mathcal{F}(\sigma_{\rho(n)})] \ldots ].
\end{aligned}    
\end{equation}
The summation in the last row is over all permutations $\rho$ of $1,\ldots,n$. 
The parameter $d_\rho$ is defined as the number of descents of the permutation $\rho$, i.e.~the number of $i$'s for which $\rho(i) > \rho(i+1)$.
Overall, for $\tilde{H}_n$, we expect nested commutators of order $n-1$ of $\mathcal{F}$ to appear. 

We now derive a closed-form formula for the effective Hamiltonian based on the above; note that Eq.~\eqref{eq:magnus_for_Htilde} can be rewritten as:
\begin{align}
\begin{split}
    &\tilde{H} = H + 
    \frac{1}{s}\int_\sigma \; \mathcal{E}(\sigma) \; \\
    & + \frac{1}{2s} \int_{\sigma_1\sigma_2}
    \left( [H, \mathcal{E}(\sigma_2) ] +  [\mathcal{E}(\sigma_1), H ] + [\mathcal{E}(\sigma_1), \mathcal{E}(\sigma_2) ]  \right) \\
    &+ \ldots \ .
\end{split}
\end{align}
As we are interested in calculating the error in QPE outcomes to lowest orders in $s$, we need the following observations:
\begin{enumerate}
    \item Based on the above Magnus expansion, we can see that for an order $p$ product formula, $\tilde{H} - H = \mathcal{O}(s^p)$. This is because the lowest contribution comes from the first term above $\frac{1}{s}\int_0^s d\sigma \mathcal{E}(\sigma)$, where we recall $\mathcal{E}(\sigma)=\mathcal{O}(\sigma^p)$.
    \item For $\tilde{H}_2$, we note that the term $\frac{1}{2s} \int_0^s d\sigma_1 \int_0^{\sigma_1} d\sigma_2 \; [\mathcal{E}(\sigma_1), \mathcal{E}(\sigma_2) ]$ will result in a contribution of order $\mathcal{O}(s^{2p+1})$, while the other two remaining terms will be of order $\mathcal{O}(s^{p+1})$.
    \item Leveraging the task in hand, for the calculation of the expectation value in Eq.~\eqref{eq:energy_shift_pert} up to (but not including) order $s^{2p+1}$, only the term $\frac{1}{s} \int_0^s d\sigma \;\langle \psi |  \mathcal{E}(\sigma)|\psi\rangle$ contributes. This is because we either have (i) more than one $\mathcal{E}$ in the nested commutators, in which case the power of $s$ in the term will be larger than or equal to $2p+1$, like in the previous observation, or (ii) we have only one $\mathcal{E}$ in the expansion, similar to e.g.~$\frac{1}{2s} \int d\sigma_1 d\sigma_2 \; [H, \mathcal{E}(\sigma_2) ]$, and the operator appearing in the outermost nested commutator should be $H$. But this makes the expectation value $\langle \psi | [H,\cdot]|\psi\rangle$ vanish, as $|\psi\rangle$ is an eigenstate of $H$.

    \item The above observations together show that for an order $p$ product formula, the QPE error can be given by:
    \begin{equation}\label{eq:qpe_error_perturbative_scriptE}
        \delta E =  \frac{1}{s} \int_0^s d\sigma \; \langle \psi |\mathcal{E}(\sigma)|\psi\rangle + \mathcal{O}(s^{2p}),
    \end{equation}
    where the $\mathcal{O}(s^{2p})$ arises due to the second order of perturbation theory in $\tilde{H}-H$ (the effect of higher order terms in Eq.~\eqref{eq:energy_shift_pert}).
\end{enumerate}
This is an important observation, as we know the form of $\mathcal{E}$ in terms of nested commutators and as a result we can calculate the QPE error up to order $s^{2p}$ using its expectation value.

We note that the above analysis is independent of the initial state $\ket{\psi_{\text{in}}}$ for the energy estimation routine. For the particular case of QPE, the only practical requirement is that it has an appreciable overlap with the target state so that the sampling happens efficiently. In the above, only the properties of the target eigenstate determine the error in obtaining the QPE outcome.

Let us now check that the first order result of \cite{yi2022spectral} is recovered in our analysis, showing that $s = \mathcal{O}(\epsilon^{1/2})$ suffices. As they study a Hamiltonian $H=A+B$, the only term appearing in $\mathcal{E}$ for first order in $s$ is $[A,B]$, which can also be written as $[H,B]$. Since $H$ appears in the commutator, the first order contribution to the QPE error will be zero, making the second order the lowest order contribution, thus $s=\mathcal{O}(\epsilon^{1/2})$.

Let us mention that in the above, the finite error of the QPE routine has not been considered, however, including it is simple and done in the supplementary material. Furthermore, we note that extending Eq.~\eqref{eq:qpe_error_perturbative_scriptE} to the case of QPE error for a degenerate subspace using degenerate perturbation theory can be done straightforwardly.

\begin{table}[!t]
    \centering
    \begin{tabular}{|c|c|c|c|}
    \hline
          & 2nd order Trotter-Suzuki & $ \qquad \mathcal{W}_2(s) \qquad $ & $\mathcal{W}_2(s/2)\mathcal{W}'_2(s/2)$ \\
         \hline
         $s$ & $\mathcal{O}(\epsilon^{1/2})$ & $\mathcal{O}(\epsilon^{1/3})$ &  $\mathcal{O}(\epsilon^{1/4})$  \\
         \hline
    \end{tabular}
    \caption{Comparing the scaling of the Trotter step $s$ as a function of the error $\epsilon$ in energy obtained through QPE, between the following three cases: 2nd order Trotter Suzuki formula (previously known) and the product formulas introduced here: $W_2(s)$ and $\mathcal{W}_2(s/2)\mathcal{W}'_2(s/2)$.}
    \label{tab:comparing_results}
\end{table}

\section{Customized Product Formulas}
As noted previously, when a nested commutator in the expansion of $\mathcal{E}$ has $H$ on its outermost commutator, it will have no contribution to the QPE error. This criterion guides us in our design of new product formulas whose QPE error have a better scaling than generic product formulas of the same order, e.g.~Trotter-Suzuki.
More concretely, for a general order $p$ product formula, the step size needs to scale as $\epsilon^{1/p}$; however, using the above prescription, if for a product formula of order $p$ all nested commutators of orders $p'$ and less, with $p \leq p'< 2p$, contain $H$ in the outermost commutator, the step size will scale as $\epsilon^{1/(p'+1)}$. This can result in an up to a quadratic improvement provided enough conditions like the above are satisfied.

\begin{figure*}[!t]
    \includegraphics[width=.67\columnwidth]{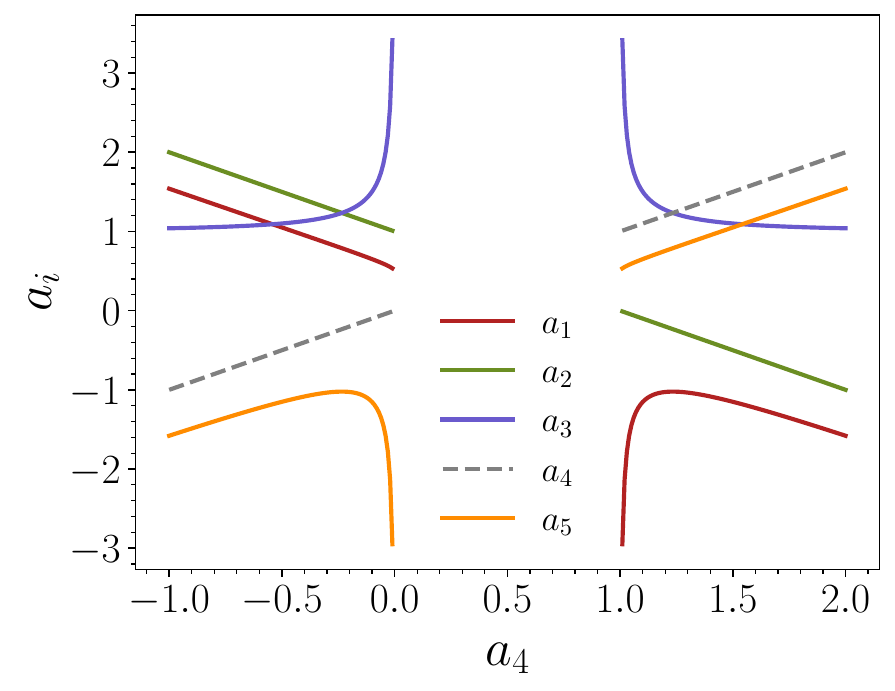}
    \includegraphics[width=.69\columnwidth]{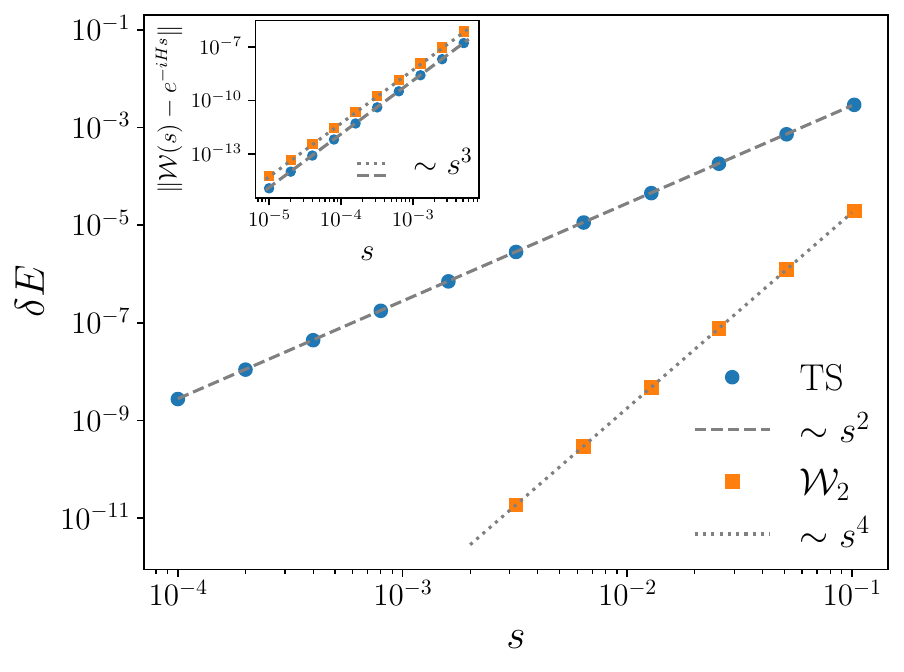}    \includegraphics[width=.69\columnwidth]{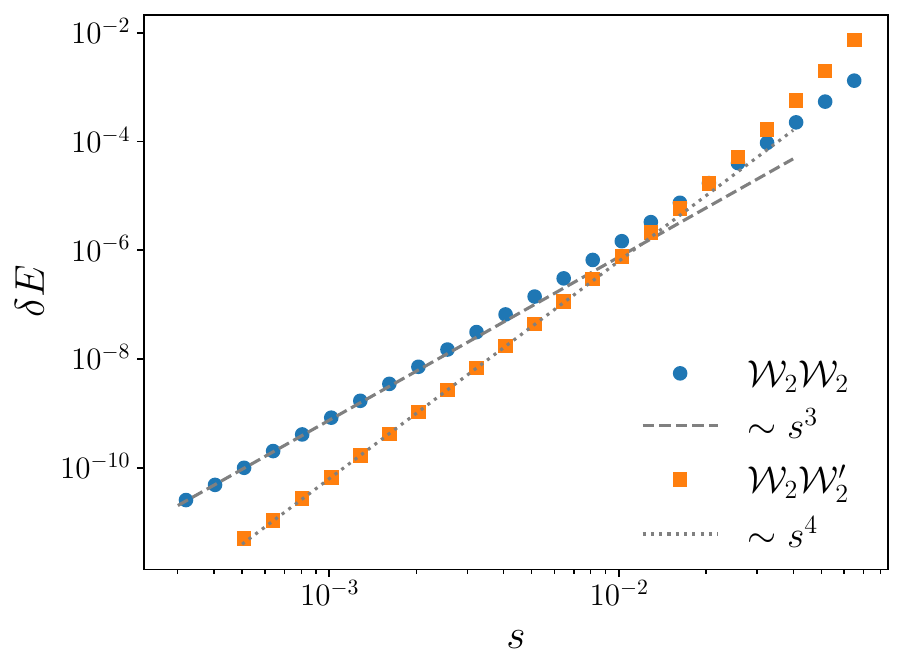}
    \caption{Left: Solutions of $a_i$'s in Eq.~\eqref{eq:prodcut_formula_5_terms}. The parameter $a_4$ is the free variable which can take real values outside of $[0,1]$.
    Middle: Comparing $\mathcal{W}_2$ and second-order Trotter-Suzuki (TS in the plot legend): energy estimation error ($\delta E$) for a Trotter step of length $s$ for an XY model Hamiltonian on a 2D $3\times 4$ lattice. We see clearly that $\mathcal{W}_2$ outperforms with $s^4$ error scaling compared to $s^2$. Inset confirms that both spectral errors scale as $s^3$. 
    Right: Comparing $\mathcal{W}_2\mathcal{W}_2$ with $\mathcal{W}_2\mathcal{W}_2'$ energy estimation error ($\delta E$) for a random $9$ qubit Hamiltonian. We see the promised $s^4$ scaling of $\mathcal{W}_2\mathcal{W}_2'$, which is better at small time-steps than that of $\mathcal{W}_2\mathcal{W}_2$. Around $s=10^{-2}$, both of the curves start deviating from their lowest order behavior due to higher order contributions showing up.}
\label{fig:plots}
\end{figure*}
As an application of this criterion, we introduce a second-order product formula that extends the result in  Ref.~\cite{yi2022spectral}.
For $H=A+B$, where $A,B$ are fast-forwardable, we claim that there exists a family of second-order product formulas
\begin{equation}\label{eq:prodcut_formula_5_terms}
    \mathcal{W}_2(s) = e^{-i A a_1 s} e^{-i B a_2 s} e^{-i A a_3 s} e^{-i B a_4 s} e^{-i A a_5 s},
\end{equation}
whose error in a single step takes the form:
\begin{equation}\label{eq:claim}
    \mathcal{W}_2(s) - e^{-iHs} = \alpha_3 \; s^3 \;  [H,[H,B]] + \mathcal{O}(s^4),
\end{equation}
with $\alpha_3$ a constant. Notice that the above can be equivalently stated in terms of the exponent operator taking the form: $\mathcal{E}(\sigma) = 3\alpha_3 \sigma^2 [H,[H,B]] + \mathcal{O}(\sigma^3)$. 
This means that using Eq.~\eqref{eq:qpe_error_perturbative_scriptE}, the shift in energy in perturbation theory will be $\mathcal{O}(s^3)$, this is contrary to the $\mathcal{O}(s^2)$ scaling for a general second order product formula.

For the claim to hold we need a few equations that the $a_i$'s must satisfy (details in \cite{SM}). This can be done with one of the $a_i$ values kept as a free variable, which we have taken as $a_4$, with the rest of the $a_i$'s plotted in terms of that, as shown in Fig.~\ref{fig:plots} (left). However, one must choose the free variable outside of the range $0<a_4<1$, otherwise it is not possible to satisfy all the conditions. Furthermore, the number of conditions involved also imply that a minimum of 5 individual exponentials is required for all of this to be satisfied.

Furthermore, if the order of the terms in $\mathcal{W}_2$ is swapped to define $\mathcal{W}_2'$ as
\begin{equation}\label{eq:w_prime_def}
    \mathcal{W}_2'(s) = e^{-i A a_5 s} e^{-i B a_4 s} e^{-i A a_3 s} e^{-i B a_2 s} e^{-i A a_1 s},
\end{equation}
the combination $\mathcal{W}_2(s/2)\mathcal{W}_2'(s/2)$, by inspection, leads to a QPE error of $\mathcal{O}(s^4)$. This can be verified by checking that assuming a solution $\mathcal{W}_2(s)$ to~\eqref{eq:claim}, we have 
\begin{align}
    &\mathcal{W}_2(s/2)\mathcal{W}_2'(s/2) - e^{-iHs} = \nonumber \\
    &\alpha_3s^3[H,[H,B]] + \alpha_4s^4[H,[H,[H,B]]] + \mathcal{O}(s^5).
\end{align}
This combination is motivated by the symmetrization of product formulas in the style of Trotter-Suzuki hierarchy, which are symmetric by construction.  Therefore, we have discovered a 9-term second order product formula with an error in energy estimation that is of the same order in $s$ as a Trotter-Suzuki fourth order product formula which itself uses 11 terms. We recap our results and that of \cite{yi2022spectral} in Table~\ref{tab:comparing_results}. Here, the scaling with the ultimate error is considered as it most manifestly shows the gain in the new product formulas.

\medskip
\section{Numerical Investigation}
We verify our findings for a $3\times 4$ square lattice XY model
    $H = -\frac{1}{2}  \sum_{\langle i j\rangle} \left( J_x  X_i X_j   + J_y Y_i  Y_j\right)$
where $\langle i j\rangle$ denotes summation over neighboring sites on the lattice. For our simulations, we set $J_x = 0.25, J_y = 0.75$, with $A$ and $B$ in the product formula set as the $X$ and $Y$ interaction terms. 

For a single Trotter step, Fig.~\ref{fig:plots} (middle) plots the comparison between the QPE error scaling of $\mathcal{W}_2(s)$ (with $a_4=-0.3$) and the 2nd-order Trotter-Suzuki for the lowest eigenvalue, where a $s^4$ vs $s^2$ order behavior is seen. The inset verifies that both product formulas are indeed of second order. 
We expect an $s^3$ behavior for $W_2$, but the $s^4$ behavior appearing here is likely due to the structure of the model we have considered. 
As a result in the following, we consider a random Hamiltonian for the task of comparing
$\mathcal{W}_2(s/2)\mathcal{W}_2'(s/2)$ with $\mathcal{W}_2(s/2)\mathcal{W}_2(s/2)$; we have chosen two Trotter steps for $W_2$ each of length $s/2$ to take an equal number of exponentials into account for both of the formulas. Therefore, in Fig.~\ref{fig:plots} (right), we compare the QPE error of $\mathcal{W}_2(s/2)\mathcal{W}_2(s/2)$ with that of $\mathcal{W}_2(s/2)\mathcal{W}_2'(s/2)$ for the lowest eigenvalue, and confirm our theory. 
The random Hamiltonian is created by summing two terms each consisting of 50 random Pauli strings on 9 qubits with random coefficients and lengths. Notice that for both product formulas, for larger $s$ values, higher order contributions start to take over.

\section{Improved Energy Estimation Error Bound for Low-Energy States}
We would like to study the energy estimation error of low-lying eigenstates of a set of Hamiltonians discussed in \cite{hejazi2024better}, wherein a better bound for the cost was provided for low-energy dynamics. This set is given by $k$-local Hamiltonians $H=\sum_{m=1}^M H_m$ acting on $N$ qubits with $M=\mathcal{O}(1)$, where $H_m$ are fast-forwardable positive semidefinite terms written as a sum of $\mathcal{O}(N)$ many $k$-local interaction terms, with each qubit appearing in at most $d$ many interactions. 
We also assume $q=\mathcal{O}(1)$ in Eq.~\eqref{eq:general_product_formula}.
For such Hamiltonians, we would like to compute the energy estimation error for low-lying states. We already know that it is evaluated by Eq.~\eqref{eq:qpe_error_perturbative_scriptE} where $\ket{\psi} = \ket{\psi_{\Delta}}$ is a low-energy eigenstate with an energy upper bounded by some value $\Delta$. 

To calculate the expectation value, we simply need to carry out an  analysis identical to the one offered in the supplementary materials of \cite{hejazi2024better} (Eqs. (45-49)) for estimating $\int_0^s d\sigma \; \|\Pi_{\le \Delta'} \mathcal{E}(\sigma) \Pi_{< \Delta}\|$, by substituting $\Delta=\Delta'$, where $\Pi_{\le \Delta}$ is the projection onto the subspace with energy at most $\Delta$. Crucially, this upper bounds any expectation value with a low-energy eigenstate whose energy is upper bounded by $\Delta$; thus we are indeed upper-bounding  $\int_0^s d\sigma \; |\langle \psi_\Delta | \mathcal{E}(\sigma) | \psi_\Delta \rangle| $. 
Furthermore, unlike the case in Eq. (45) of \cite{hejazi2024better}, we no longer have a difference $\Delta'-\Delta$ scaling with $N$. Therefore, as concluded therein, $\frac{1}{s}\int_0^s d\sigma \; |\langle \psi_\Delta | \mathcal{E}(\sigma) | \psi_\Delta \rangle| $ would scale as $\mathcal{O}(\Delta_f^{p+1} s^{p})$ where $\Delta_f -\Delta = \mathcal{O}(\log(N)^{p+1})$.
 
The same conclusion holds for all higher order contributions up to $s^{2p-1}$; thus according to the above analysis, and assuming $\Delta$ has a slow enough dependence on $N$ (like the case of the ground state), for orders $p$ to $2p-1$  the spectral-based bound $\text{poly}(N)$ used for bounding the nested commutators can be improved exponentially. However, from $s^{2p}$ onward, without any additional information, the terms can only be bounded by polynomials in $N$, especially since higher order contributions in $\tilde{H} - H$ need to be considered as well.
Thus taking all of this into account, the time step can have a scaling as:
$s= \mathcal{O}\left(\frac{\epsilon^{1/p} }{\log(N)^{1+1/p}}+ \frac{\epsilon^{1/(2p)} }{\text{poly(N)}}\right)$. 

Assuming the polylogarithmic contributions in $N$ are subdominant compared with the polynomial ones, this shows a quadratic improvement in scaling with $\epsilon$.

\section{Discussion}
In this work, we extend the previous perturbation theory based analysis \cite{yi2022spectral} of quantum phase estimation error of product formulas, and compute a new closed form formula by employing the Magnus expansion. We apply this approach to devise a new method of product formula constructions geared toward energy estimation errors instead of spectral errors.
As an application, we consider a class of Hamiltonians decomposing to two fast-forwardable terms, which includes electronic structure Hamiltonians, and we find custom second-order product formulas which achieve cubic and quartic energy estimation error scalings with respect to step-size $s$. This generalizes the result of Ref.~\cite{yi2022spectral} to second order and we demonstrate, theoretically and numerically, that this improves upon the standard second-order Trotter-Suzuki product formula. 
Lastly, we combine our Magnus-based analysis with the results in low energy dynamics of Ref.~\cite{hejazi2024better}. For the task of energy estimation of the low-lying eigenstates of a set of local Hamiltonians, we show that a quadratic improvement can be obtained when compared with vanilla spectral-based bounds. 

Although the latter set is limited in that the Hamiltonian must decompose to positive semidefinite terms that are fast-forwardable, we conjecture that even with relaxing this positivity assumption, obtaining better bounds with a similar, but improved, analysis should be possible (see also the discussion in \cite{hejazi2024better}). 
Similarly, designing custom product formulas achieving better energy estimation error for more general Hamiltonians and higher orders is left to future studies.

\bibliography{main.bib}

\newpage

\onecolumngrid

\bigskip
~
\bigskip

    \centerline{ {\huge {\bf Supplementary material}} }

\bigskip

\section{Equations for the custom product formula coefficients}
To determine the custom product formula coefficients $a_i$'s, we will need the expression of $\mathcal{F}(s)$ in terms of nested commutators (\cite{childs2021theory}):
\begin{equation}\label{eq:expansion_of_F_nested}
\begin{aligned}
    \mathcal{F}(s) &= H \\
    &+ (-is) \sum_{\nu=1}^q \sum_{\nu_1=\nu+1}^q \tilde{f}_{\nu_1,\nu} [H_{\nu_1},H_\nu]\\
    &+ (-is)^2 \sum_{\nu=1}^q \sum_{\nu_1=\nu+1}^q \sum_{\nu_2=\nu_1}^q \tilde{f}_{\nu_2\nu_1,\nu} [H_{\nu_2},[H_{\nu_1},H_\nu]]\\
    &+ (-is)^3 \sum_{\nu=1}^q \sum_{\nu_1=\nu+1}^q \sum_{\nu_2=\nu_1}^q \sum_{\nu_3=\nu_2}^q \tilde{f}_{\nu_3\nu_2\nu_1,\nu} [H_{\nu_3},[H_{\nu_2},[H_{\nu_1},H_\nu]]]\\
    &+\ldots ,
\end{aligned}
\end{equation}
where the index $\nu_i$ shows the Hamiltonian term that goes into the position $\nu_i$ of the product formula;
for example, for the custom-designed formula introduced in the main text $\mathcal{W}_2(s)$, we have $q=5$ and $H_{\nu_i}$ is equal to $A$ for odd $\nu_i$ and equal to $B$ for even $\nu_i$. The coefficients can be calculated as:
\begin{equation}
\begin{aligned}\label{eq:f_nu_eqs}
    \tilde{f}_{\nu_1,\nu} = a_\nu a_{\nu_1}; \qquad &\nu_1> \nu, \\
    \tilde{f}_{\nu_2\nu_1,\nu} = a_\nu a_{\nu_1} a_{\nu_2}; \qquad &\nu_2 > \nu_1> \nu, \qquad \tilde{f}_{\nu_1\nu_1,\nu} = \frac{1}{2!}  a_\nu a_{\nu_1}^2,\\
    \tilde{f}_{\nu_3\nu_2\nu_1,\nu} = a_\nu a_{\nu_1} a_{\nu_2} a_{\nu_3}; \qquad &\nu_3 > \nu_2 > \nu_1> \nu, \qquad \tilde{f}_{\nu_1\nu_1\nu_1,\nu} = \frac{1}{3!}a_\nu a_{\nu_1}^3, \qquad \ldots\\
    \vdots
\end{aligned}
\end{equation}
In the main text we have used an alternative form in which the indices of the coefficients $f_{\{m_i\},m}$ run over Hamiltonian terms and not the product formula terms (which are also Hamiltonian terms except that they include repetitions). Note that the above expansion can also be transformed into such a form by summing terms that are identical in the expansion. In the following, we switch back to that representation.

We can now identify the equations that our custom product formula must satisfy. We recall our claim: There exist a second order product formula for $H=A+B$, whose error in a single Trotter step reads:
\begin{equation}\label{eq:claim2}
    \mathcal{W}_2(s) - e^{-iHs} = \alpha_3 \; s^3 \;  [H,[H,B]] + \mathcal{O}(s^4),
\end{equation}
with $\alpha_3$ a constant and with $\mathcal{W}_2$ defined as:
\begin{equation}\label{eq:prodcut_formula_5_terms2}
    \mathcal{W}_2(s) = e^{-i A a_1 s} e^{-i B a_2 s} e^{-i A a_3 s} e^{-i B a_4 s} e^{-i A a_5 s}.
\end{equation}

The above claim can also be equivalently stated in the following way. Expanding $\mathcal{W}_2$ as:
\begin{equation}
    \mathcal{W}_2(s) = \mathcal{T} e^{-i \int_0^s d\sigma \mathcal{F}(\sigma)},
\end{equation}
with:
\begin{equation}\label{eq:F_5_terms}
\begin{aligned}
    \mathcal{F}(\sigma) &= \left( f_A A + f_B B \right) \\
    &+ \sigma  f_{A,B} [A,B]  \\
    &+ \sigma^2 \left( f_{AA,B} [A,[A,B]] + f_{BA,B} [B,[A,B]] \right) \\
    &+  \sigma^3 \Big( f_{AAA,B} [A,[A,[A,B]]] + f_{ABA,B} [A,[B,[A,B]]]  \\
    &\quad + f_{BAA,B} [B,[A,[A,B]]] + f_{BBA,B} [B,[B,[A,B]]] \Big) \\
    &+\mathcal{O}(\sigma^4),
\end{aligned}    
\end{equation} 
we note that Eq.~\eqref{eq:claim2} is equivalent to $\mathcal{F}(\sigma)$ taking the form $3\alpha_3 \sigma^2 [H,[H,B]]$ to lowest order in $\sigma$. 
For the claim to hold, we use Eq.~\ref{eq:f_nu_eqs} to obtain the equations that need to be satisfied:
\begin{itemize}
    \item The first order in $s$ in Eq.~\eqref{eq:F_5_terms}, should be $H$ and thus $f_A=f_B=1$. Equivalently:
    \begin{equation}\label{eq:a_eqs_1}
    \begin{aligned}
        a_1 + a_3 + a_5 &= 1, \\
        a_2 + a_4 &= 1 .
    \end{aligned}
    \end{equation}

    \item The second order in $s$ in Eq.~\eqref{eq:F_5_terms} should vanish, resulting in:
    \begin{equation}\label{eq:a_eqs_2}
    \begin{aligned}
        a_1 (a_2+a_4) - a_2 (a_3+a_5) + a_3 a_4 - a_4 a_5 = 0.
    \end{aligned}
    \end{equation}

    \item The third order terms in $s$ in Eq.~\eqref{eq:F_5_terms} should have equal coefficients, resulting in:
    \begin{equation}\label{eq:a_eqs_3}
    \begin{aligned}
        \frac{1}{2}a_1^2(a_2+a_4) - a_1 a_2 (a_3+a_5) - a_1 a_3 a_4 + a_1 a_4 a_5 \\
        + \frac12 a_3^2 a_4 - a_3 a_4 a_5 \qquad \\
        = -\frac12 a_2^2 (a_3+a_5) + a_2 a_3 a_4 - a_2 a_4 a_5 - \frac{1}{2} a_4^2 a_5.
    \end{aligned}
    \end{equation}
\end{itemize}

\section{Details of the perturbation theory analysis}
Herein, we build on the detailed perturbation theory analysis in Ref.~\cite{yi2022spectral}. Recall $\tilde{H}(s) | \psi(s) \rangle = \tilde{E}(s) | \psi(s) \rangle$, where $\ket{\psi(0)}=\ket{\psi}$ and $\ket{\psi}$ is the eigenstate of $H$ with eigenvalue $E$. As in the main text, we drop the $s$ dependence of $\tilde{H},\tilde{E}$ for convenience. We expand $| \psi(s) \rangle$ in a perturbative fashion as:
\begin{equation}
    | \psi(s) \rangle = \sqrt{1- f(s)^2} | \psi \rangle + f(s) | \phi(s) \rangle,
\end{equation}
with $f(s) > 0$ and $\langle \psi | \phi(s) \rangle = 0 $.  
Defining $V = \tilde{H} - H$, we note that $V = \mathcal{O} \left( s^p  \alpha_{\text{comm},p}  \right)$. $\alpha_{\text{comm},p} $ is defined as the sum of the norms of all the nested commutators of order $p$ with the same coefficients that appear in the expansion of $\mathcal{E}$. Considering the projector $P(s) = | \psi(s) \rangle \langle \psi(s) |$, we have the following two relations:
\begin{equation}\label{eq:proj_in_termsof_f}
    \| P(s) - P(0) \|^2 = f(s)^2,
\end{equation}
\begin{equation}\label{eq:deriv_proj}
    \left\| \frac{d}{ds} P(s) \right\| \leq \frac{\| \frac{d}{ds} \tilde{H}(s)\|}{\lambda}.
\end{equation}
The second equation is taken from Lemma 8 of \cite{jansen2007bounds}, where $\lambda$ is the spectral gap between the eigenvalue of interest and other eigenvalues. Integrating Eq.~\eqref{eq:deriv_proj}, we obtain:
\begin{equation}
    \| P(s) - P(0) \| \leq s \max_\sigma
    \frac{1}{\lambda} 
    \left\| \frac{d}{d\sigma} \tilde{H}(\sigma) \right\|
\end{equation}
Combining this with Eq.~\eqref{eq:proj_in_termsof_f} gives:
\begin{equation}
    f(s) = \mathcal{O} \left( s^p  \alpha_{\text{comm},p} / \lambda \right)
\end{equation}
where we have used that the lowest order contribution to $\tilde{H}$ is obtained from $\tilde{H}_1$. Now, we can expand the exact energy difference as:
\begin{equation}
\begin{aligned}
    \delta E & =\langle \psi(s) |\tilde{H} | \psi(s) \rangle - \langle \psi |H | \psi \rangle \\
    &= - \langle \psi |H | \psi \rangle f(s)^2 +  \langle \phi(s) | H | \phi(s) \rangle f(s)^2 \\
    &+ \langle \psi | V | \psi \rangle (1-f(s)^2 ) \\
    &+ 2\text{Re}(\langle \psi | V | \phi(s) \rangle) \, f(s)\sqrt{1-f(s)^2 } \\
    &+ \langle \phi(s) | V | \phi(s)\rangle f(s)^2.
\end{aligned}    
\end{equation}
In general, the term $\langle \psi | V | \psi\rangle$ gives the lowest order contribution in $s$, and in general for an order $p$ product formula, this gives a contribution $\mathcal{O}(s^p)$. More precisely, we have:
\begin{equation}
\begin{aligned}
    \delta E &= \mathcal{O} \Big (\frac{s^{2p}}{\lambda^2} \alpha_{\text{comm},p}^2 \| H \| \\
    &+ s^{p'} \alpha_{\text{comm},p'} \\
    &+ \frac{s^{2p}}{\lambda} \alpha_{\text{comm},p}^2 \\
    &+ \frac{s^{3p}}{\lambda^2} \alpha_{\text{comm},p}^3\Big),
\end{aligned}    
\end{equation}
where $p' \geq p$ in the above equation is the first order in $s$ for which $\langle \psi | V | \psi\rangle$ is nonzero. Given the main result of this work for custom designed product formulas, the expectation value $\langle \psi | V | \psi \rangle$ can be of higher order, i.e. $p' > p$. For example, for $p=2$ and $\mathcal{W}_2$ defined in Eq.~\eqref{eq:prodcut_formula_5_terms2}, $\langle \psi | V | \psi \rangle$ is $\mathcal{O}(s^3)$ and for $\mathcal{W}_2(s/2) \mathcal{W}_2'(s/2)$, it is $\mathcal{O}(s^4)$.  

Lastly, let us make a remark on the error incurred by QPE or the energy estimation routine, where there is also an error $\varepsilon$ in estimating $\tilde{E}$ itself, scaling the cost of the algorithm by $\Omega(1/\varepsilon)$. For a total target error $\epsilon$, we must satisfy $\delta E  + \varepsilon \le \epsilon$. Therefore, when $\varepsilon$ is already chosen, $s$ must be small enough such that $\delta E \le \epsilon - \varepsilon$. Alternatively, one may attempt to optimize the total cost of the algorithm by targeting $\varepsilon = \alpha \epsilon, \ \delta E = \beta \epsilon$ and optimizing the cost under the constraint $\alpha+\beta=1$. Either way, our  analysis shows that the scaling of the cost by $\epsilon$ due to our custom product formulas is more efficient, i.e. $\epsilon^{-1/3}$ or $\epsilon^{-1/4}$ instead of $\epsilon^{-1/2}$. 

Notice that the discussion above generalizes to any energy estimation routine incurring an error $\varepsilon$. The purpose of our work is to show the user how small $s$ must be so that, taking into account the error $\varepsilon$ by the energy estimation, they satisfy their target accuracy $\epsilon-\varepsilon$.

\end{document}